\documentclass[epj]{webofc}
\usepackage[varg]{txfonts}   % Web of Conferences font
\newcommand{\eqb}{\begin{equation}}
\newcommand{\eqe}{\end{equation}}
\newcommand{\dmb}{\begin{displaymath}}
\newcommand{\dme}{\end{displaymath}}
\newcommand{\pd}{\partial}

\newcommand{\eab}{\begin{eqnarray}}
\newcommand{\eae}{\end{eqnarray}}

\wocname{EPJ Web of Conferences}
\woctitle{ICNFP 2018}
\begin{document}
\selectlanguage{english}
\title{SU(2) Quantum Yang-Mills Thermodynamics:\\ Some Theory and Some Applications}
%
% subtitle (optional, strongly discouraged)

\author{Ralf Hofmann\inst{1}\fnsep\thanks{\email{r.hofmann@thphys.uni-heidelberg.de}} 
        % etc.
}

\institute{Institut f\"ur Theoretische Physik, Universit\"at Heidelberg, Philosophenweg 16, 
D-69120 Heidelberg, Germany}

\abstract{%
In the first part of this talk we review some prerequisites for and 
essential arguments involved in the construction of 
the thermal-ground-state estimate underlying the deconfining phase in the thermodynamics of 
SU(2) Quantum Yang-Mills theory and how this structure supports its 
distinct excitations. The second part applies deconfining SU(2) Yang-Mills thermodynamics to the 
Cosmic Microwave Background in view of (i) a modified temperature-redshift 
relation with an interesting link to correlation-length criticality 
in the 3D Ising model, (ii) the implied minimal 
changes in the dark sector of the cosmological model, and (iii) 
best-fit parameter values of this model when confronted with the spectra of the 
angular two-point functions TT, TE, and EE, excluding the low-$l$ physics. The latter, which so far is treated in an incomplete way due to the omission of radiative effects,  
is addressed in passing.      
}
\maketitle
\section{Introduction}
\label{intro}

Based on perturbative asymptotic freedom at zero temperature \cite{GrossWilczek,Politzer,KryplovichI} the predictions of thermodynamical potentials and plasma polarisation effects in four-dimensional Yang-Mills thermodynamics suggest themselves (naively as it turns out) to small-coupling expansions at high temperature, see, e.g. \cite{BraatenPisarskiI}. That such an approach does not lend itself to the usual, 
physically approximating, truncated asymptotic series in powers of a small fundamental 
coupling $g$ was pointed out early in \cite{Linde}, however. In the talk by Thierry Grandou within this session \cite{Grandou2018} this is 
demonstrated exemplarily for the damping rate of a heavy fermion propagating through a hot gauge 
plasma \cite{BraatenPisarskiII}, suggesting the existence of an infinite and thus hardly 
controllable tower of momentum scales, each implying a resummation within an 
associated class of loop diagrams. On the other hand, the approach of lattice gauge 
theory to the computation of thermodynamical quantities, albeit genuinely nonperturbative 
\cite{Deng,Bielefeld}, is subject to infrared artefacts due to 
finite-size limitations \cite{Karsch_finite_size_scaling}, the imprecisely known 
nonperturbative beta function \cite{Deng}, and the 
prescription of seemingly unphysical boundary conditions in the so-called 
integral method \cite{Bielefeld}. 

All this has nurtured the suspicion that deconfining Yang-Mills thermodynamics 
actually is subject to highly nonperturbative physics, even at high 
temperatures. Talking of nonperturbative field configurations, the 1970ies and early 1980ies have seen tremendously fertile developments in constructing, in principle, all exact, 
(anti)selfdual solutions to the Euclidean 
Yang-Mills equation on ${\bf R}^4$ \cite{BPST,ADHM} and on the 
four-torus ${\bf T}^4$ \cite{Nahm}. Note that such (classical) configurations imply the weight $\exp\left(-8\pi^2 |k|/g^2\right)$ in the partition 
function, where $k$ refers to the topological charge, with the perturbative expansion point 
$g=0$ being an essential zero (coefficients of the 
expansion of such a weight into powers of the coupling $g$ are identically zero to all orders). Therefore, (anti)selfdual field configurations are ignored in perturbation theory.

A genuine shortcoming of the perturbative treatment in Quantum Yang-Mills theory 
is the triviality of its a priori ground-state estimate, implying the need for renormalization: the mathematical meaningless situation, arising from unconstrained quantum fluctuations (Paul A. M. Dirac), is remedied by subtractions of infinities and re-definitions of the parameters in the defining action. 
A major accomplishment in the development of perturbatively accessed gauge theory as an essential 
pillar of the successful Standard Model of Particle Physics (SMPP) was 
to show that this renormalization programme does not spoil the predictivity of the theory in keeping untouched the set of parameters that characterises the defining Lagrangian \cite{'tHooft,Veltman}. 
Driven by the short-range nature of the weak interactions, gauge-symmetry breaking was introduced 
into the SMPP in the framework of the Glashow-Salam-Weinberg theory in terms of an SU(2) fundamentally charged scalar (Higgs) field subject to a renormalisable selfinteraction. Such a Higgs sector, however, only vaguely represents the phase structure and associated gauge-symmetry breaking patterns of the pure and nonperturbatively approached Yang-Mills theory \cite{HofmannBook}. The nonperturbative a priori estimate of each phase's ground state in the latter case clearly links to its (gapped or ungapped) excitations, thereby ruling both classical and quantum radiative behaviour in distinct spectral domains \cite{HofmannBook}. 

Based on the classical Euclidean Yang-Mills action void of matter fields, the key to dynamical gauge-symmetry breaking in pure Quantum Yang-Mills theory is the construction of a useful thermal-ground-state estimate in the deconfining phase. Here we only review the SU(2) case, the more involved situation of SU(3) is treated in \cite{Hofmann2005,HofmannBook}. In short, this construction represents spatially densely packed 
(anti)caloron centers with overlapping peripheries in terms of an inert, adjoint scalar field $\phi$ 
and an effective pure-gauge configuration $a_\mu^{\rm gs}$, respectively, both arising upon spatial coarse-graining. The (anti)caloron field is an exact, nonpropagating solution to the (anti)selfduality or BPS equation. As a consequence, a vacuum equation of state emerges subject to an energy density $\rho^{\rm gs}=4\pi \Lambda^3T$, $T$ denoting temperature. Here, the mass scale $\Lambda$ represents a constant of integration in the potential 
$V(\phi)$ being the solution of a first-order ODE in the field $\phi$ \cite{GiacosaHofmann2007}. This ODE represents a consistency requirement for $\phi$ simultaneously solving the Euler-Lagrange and BPS equations associated with its effective action. Being charged under the adjoint representation of SU(2), $\phi$ is an element of the algebra su(2) and breaks SU(2) down to U(1). Thus there are two massive quasi-particle 
modes and one massless one in the effective gauge field. 

If the wavelength $l$ associated with the propagation of the massless mode either falls below the length scale $s=\pi T |\phi|^{-2}$ or even below the length $|\phi|^{-1}$, this disturbance either is forced into off-shellness by a violation of the Bianchi identity for the U(1) field strength tensor due to (anti)selfdual monopoles or it probes (anti)caloron centers, respectively. On the other hand, for $l\gg s$ the massless mode represents undulating and propagating repolarisations of (anti)selfdual dipole densities, that is, it propagates on-shell as an electromagnetic wave \cite{HofmannGrandou2015}. Probing (anti)caloron centers indeterministically provokes  fluctuations, each carrying quanta of energy and momentum $\hbar k$ where $k\equiv 2\pi/l$ \cite{Hofmann2016Entr}. (From now on, we set the speed of light $c$ in vacuum and $\hbar$ equal to unity: $c=\hbar=1$.) This is because the only physically continuable quantity from imaginary to real time, which can be provided by an (anti)caloron center, is its Euclidean action. Notice that this action, when interpreted as a winding number, is localized at the (anti)caloron's inmost spacetime point and, in matching with the effective theory, must be interpreted as Planck's quantum of action $\hbar$ \cite{KavianiHofmann2012}. (In association with the physics of (anti)caloron centers only static quantities have an interpretation in Minkowskian spacetime, the Euclidean time dependence of, e.g., the (anti)caloron field strength is void of meaning.) Massive modes always represent the quantum physics of (anti)caloron centers regardless of their wavelength and thus associate with quanta of energy $\sqrt{4e^2|\phi|^2+k^2}$ and momentum $k$, respectively. Here $e$ denotes the coupling in the effective theory for the deconfining phase whose $T$ dependence is governed by the Legendre transformation between pressure $P$ and energy density $\rho$ on the one-loop level \cite{Hofmann2005}. There is a singled-out, physical, and completely fixed gauge in the 
deconfining phase (unitary-Coulomb) in which the kinematic constraints, arising from the demand that 
radiative corrections and/or 4-vertices must not probe (anti)caloron centers (to allow for this would amount to a double counting of thermal quantum effects because these are already included in the propagators), 
enjoy a simple formulation. As a consequence, thermodynamical quantities are computable with 99\%-accuracy in only considering the above sketched a priori estimate of the thermal ground state plus free (quasi-particle) fluctuations \cite{Hofmann2005}. For an improvement of the ground-state's a priori estimate, the mild violation of spatial isotropy and homogeneity introduced by peripheral overlaps and packing voids of (anti)caloron centers, can be considered in terms of radiative corrections \cite{Schwarz2007}. 
Potentially, these are subject to resummations to all loop orders, see \cite{Bischer2017} for the case of dihedral diagrams in the massive sector.   

This talk is organised as follows. In Sec.\,\ref{sec-1} we discuss (anti)calorons with topological charge modulus one of trivial (Harrington-Shepard) and nontrivial holonomy (Lee-Lu-Kraan-van-Baal), and we review essential arguments involved in the construction of the a priori estimate of the thermal ground state. Subsequently, the ensuing ground-state structure and the properties of excitations are elucidated, including kinematic constraints governing the radiative corrections. In addition, some arguments on why a physics model describing the so-far probed electromagnetic spectrum needs to invoke a degree-of-thermalisation dependent mixing of Cartan subalgebras in at least two SU(2) Yang-Mills theories are provided. In Sec.\,\ref{sec-2} we review the implications of deconfining SU(2) Yang-Mills thermodynamics once the postulate is made that any spatially sufficiently extended 
thermal photon gas, and in particular the Cosmic Microwave Background (CMB), are subjected to 
this gauge dynamics. In such a situation, to determine the Yang-Mills scale $\Lambda$ (or critical temperature $T_c$ of the deconfining-preconfining phase transition) the isotropic cosmic radio background (unexplained extragalactic emission) and associated low-frequency excesses in line temperatures over the CMB baseline temperature 
of $T_0=2.725\,$K, see \cite{Arcade2} and references therein, are taken as an indication for the onset of the Meissner effect due to condensed (electric) monopoles. (An electric-magnetically dual interpretation of U(1) charges in SU(2) Yang-Mills theory is implied by the fine-structure constant being dimensionless 
and $e\propto \hbar^{-1/2}$ \cite{KavianiHofmann2012}.) This is because a Meissner massiveness implies the evanescence of conventionally equipartitioned electromagnetic waves in the deep Rayleigh-Jeans regime. Evanescent waves, however, do not propagate and thus can no longer be 
globally thermalised. This moves their spectral intensity away from 
that of classical thermal equipartitioning. Rather, a pile-up of low-frequency modes around 
frequency zero takes place now (the shortest-range fluctuations of lowest 
volume-integrated classical energy) whose spectral width essentially is given by the Meissner mass $m_\gamma$ \cite{Hofmann2009}. Since $m_\gamma$ is small on the scale of $T_0$ and 
since $m_\gamma$ rises with infinite slope in lowering $T$ through $T_c$ \cite{Hofmann2005} one concludes 
that $T_0=T_c$ to a high precision. This prompts the name SU(2)$_{\rm CMB}$ for the Yang-Mills theory describing extended thermal photon gases. Once $T_c$ is fixed in terms of $T_0$ the temperature-redshift ($z$) relation of SU(2)$_{\rm CMB}$ in an FLRW universe can be computed from the demand of energy conservation for this fluid. The $T$-$z$ relation for asymptotically high $z$ can be stated 
exactly as a linear relation whose coefficient coincides to $10^{-5}$ accuracy with the critical 
exponent for the correlation length of the magnetisation in the 3D Ising model. We argue that this is expected because both theories 
are in the same universality class and since a conformal map from a fictitious Yang-Mills 
temperature to the physical Ising temperature can be constructed in the 
critical regime. For the correlation length to diverge in 
the usual power-law fashion one deduces its exponentially faster divergence compared to the 
spatial system size. This implies that critical behaviour is well 
discernible on a {\sl finite}, spatially expanding system of fixed heat content. The remainder 
of Sec.\,\ref{sec-2} reviews the application of the $T$-$z$ relation of SU(2)$_{\rm CMB}$ to the cosmological model. There are simple arguments which imply a modified dark sector. We provide recently obtained (and not yet published) results for fits to the Planck data of the TT, TE, and EE angular power spectra (using a modified version of CLASS and CAMEL) and discuss the values of some of the thus extracted cosmological parameters. In particular, the value of the Hubble constant 
$H_0$ almost perfectly coincides with the central value of a recent local-cosmology 
extraction from the distance-redshift curve of SNe Ia 
\cite{Riess2018}  --  in contrast to the fits of 
the standard $\Lambda$CDM model to the same CMB data. Also, the new fits favour a low redshift 
of re-ionisation which agrees with pin-downs of the onset of the Gunn-Peterson 
trough in the spectra of high-$z$ quasars, see, e.g., \cite{Becker}. Finally, we provide an outlook on how the spectral overshoot in TT at low $l$, which is presently predicted by SU(2)$_{\rm CMB}$, 
can potentially be cured by taking screening effects into account at low $z$. 
In Sec.\,\ref{sec-3} we give a brief summary 
and discuss future research.      
  
\section{Thermal ground state and (quasi-particle) excitations }
\label{sec-1}
\subsection{Charge-modulus one (anti)calorons}
The Euclidean (second-order) Yang-Mills equations are satisfied by solutions to the first-order equations 
\eqb
\label{selfdual}
F_{\mu\nu}=\pm \frac12\epsilon_{\mu\nu\kappa\lambda}F_{\kappa\lambda}\equiv \tilde{F}_{\mu\nu}\,,
\eqe
where $F_{\mu\nu}\equiv\pd_\mu A_\nu-\pd_\nu A_\mu-ig[A_\mu,A_\nu]\equiv F^a_{\mu\nu}t^a\in \mbox{su(2)}$ 
is the field-strength tensor, $\epsilon_{\mu\nu\kappa\lambda}$ the totally antisymmetric 
symbol in 4D with $\epsilon_{4123}=1$, $g$ the fundamental gauge coupling, $t^a$ ($a=1,2,3$) the set of 
SU(2) generators in the fundamental representation, normalised as tr\,$t^a t^b=\frac12\delta^{ab}$, 
and $A_\mu\in \mbox{su(2)}$ denotes the gauge potential. In general, configurations, which obey Eq.\,(\ref{selfdual}), associate with vanishing energy-momentum. Thus they do not propagate and 
therefore qualify as ground-state inducers. Specifically, at finite temperature one 
is led to consider periodic gauge-field configurations, 
$A_\mu(x_4=0,\mathbf{x})=A_\mu(x_4=\beta,\mathbf{x})$, where $\beta\equiv 1/T$. In the 
following we set $x_4=\tau$. Since the group manifold of SU(2) is the three-sphere $S_3$ and since 
$\Pi_3(S_3)={\bf Z}$ solutions to Eq.\,(\ref{selfdual}) with finite action $S$ on ${\bf R}^4$, whose boundary is 
$\pd {\bf R}^4=S_3$, -- instantons -- are stabilized by and classified according to, e.g., their topological charge $k=\frac{1}{16\pi^2}\mbox{tr}\,\int d^4\,F_{\mu\nu}\tilde{F}_{\mu\nu}\in {\bf Z}=\pm \frac{g^2}{8\pi^2}S$. The case $k=\pm 1$ was discovered in \cite{BPST}, the associated gauge-field configuration exhibiting 
winding about $S_3$ at spacetime infinity. The construction of all instantons is subject to an 
algebraic constraint ((anti)selfduality on a point), representing data such as $k$, scale variables $\rho_i$, 
locations of centers $x_i$, and su(2) orientations thereof \cite{ADHM}. However, an important subset 
of configurations with $|k|>1$ can be obtained by virtue of the observation that a large gauge 
transformation $\Omega\in\,$SU(2) applied to the solution of \cite{BPST} dewinds it at infinity at the price of introducing winding about the point of maximum action density. This re-casts the gauge-field configuration to the following form
\eqb
\label{singgauge}
A_\mu=\tilde{\eta}^a_{\mu\nu} t_a\partial_\nu\log\Pi(\tau,r)\,,
\eqe     
where $\tilde{\eta}^a_{\mu\nu}$ is the anti-symmetric, anti-selfdual 't Hooft symbol, arising from the projection of the pure-gauge configuration $A^{\rm pg}_\mu=i\Omega\pd_\mu\Omega^\dagger$ onto 
the generators $t^a$. (The (anti)instanton solution is obtained by replacing $\tilde{\eta}^a_{\mu\nu}$ by 
its selfdual counterpart $\eta^a_{\mu\nu}$.) The important fact about Eq.\,(\ref{singgauge}) is that the prepotential $\Pi=1+\frac{\rho^2}{(x-x_0)^2}$ of an (anti)instanton of $|k|=1$ ($x_0$ denoting the spacetime point of maximum action density -- the (anti)instanton center) can be promoted to an arbitrary integer topological charge by superposition of terms $\frac{\rho_l^2}{(x-x_{l})^2}$ where $\{\rho_l\}$ and $\{x_{l}\}$ \cite{'tHooft1976,JackiwRebbi} are the associated sets of scale parameters and centers, respectively. In particular, setting $\rho_l\equiv \rho$ and $x_{l,4}=l\beta+\tau_0$, $\mathbf{x}_{l}\equiv \mathbf{x}_0$  for $-\infty\le l\le +\infty$ renders the prepotential $\Pi$ periodic in $\tau$. Without an essential restriction of generality we may specialise to $\mathbf{x}_0=\tau_0=0$ to obtain \cite{HS1977}
\eqb
\label{prepot}
\Pi=1+\frac{\pi\rho^2}{\beta r}\frac{\sinh\frac{2\pi r}{\beta}}{\cosh\frac{2\pi r}{\beta}-\cos\frac{2\pi\tau}{\beta}}\,,
\eqe
where $r\equiv |\mathbf{x}|$. On the slice $(0\le\tau\le\beta)\times{\bf R}^3$ the configuration associated 
with Eqs.\,(\ref{singgauge}) and (\ref{prepot}) possesses exactly one unit of 
topological charge localised at $x_0=0$. Exhibiting a Polyakov loop at spatial infinity of value unity, 
this configuration -- the Harrington-Shepard (HS) (anti)caloron -- is said to be of trivial holonomy.  

It is necessary for our discussion in Sec.\,\ref{sec2-2} to analyse this solution as a function of distance to its center. This was done in \cite{GPY1983} with the following result. For $|x|\ll\beta$ ($|x|\equiv\sqrt{x_\mu x_\mu}$) the prepotential takes the form 
$\Pi=\left(1+\frac{\pi}{3}\frac{s}{\beta}\right)+\frac{\rho^2}{x^2}$, where $s\equiv\frac{\pi\rho^2}{\beta}$, such that the solution $A_\mu$ represents an ordinary $k=1$ instanton with a 
rescaled scale parameter $\rho^\prime$ obeying ${\rho^\prime}^2=\frac{\rho^2}{1+\frac{\pi s}{3\beta}}$. For $r\gg\beta$ the configuration becomes {\sl static} because then $\Pi= 1+\frac{s}{r}$. Moreover, 
for $\beta\ll r\ll s$ it represents a static, selfdual monopole, $E^a_i=B_i^a\sim-\frac{\hat{x}^a\hat{x}_i}{r^2}$, and for $r\gg s$ one observes a static 
selfdual dipole, $E_i^a=B_i^a=s\frac{\delta_i^a-3\,\hat{x}^a \hat{x}_i}{r^3}$ of dipole moment $s$ where 
$\hat{x}\equiv\frac{\mathbf{x}}{|\mathbf{x}|}$.  

The $|k|=1$ solutions of nontrivial holonomy were constructed \cite{KraanVanBaal,LeeLu} based on Werner 
Nahm's transformation between (anti)selfdual Yang-Mills fields on ${\bf T}^4$ and 
the dual torus $\tilde{\bf T}^4$ \cite{Nahm}. The dual torus 
to $S_1\times {\bf R}^3$ is $S_1\times{\mathbf{0}}$ such that selfdual data need to be 
prescribed on a real interval only to construct the selfdual gauge field on $S_1\times {\bf R}^3$ via the Nahm transformation. Thus, instead of a 4D PDE only an ODE 
subject to certain jump conditions needs to be solved which reflect the demand for overall magnetic 
charge neutrality of a system of BPS monopole and its antimonopole (magnetic charge w.r.t. U(1)$\subset$SU(2) surviving the gauge-symmetry breaking imposed 
by $A_4(\tau,|\mathbf{x}|\to\infty)\not=0$). The thus obtained nontrivial-holonomy configuration 
can be connected smoothly to the above-discussed trivial-holonomy (anti)caloron. 
Note that the trivial-holonomy limit transmutes a pair of localised magnetic BPS monopole and 
its antimonopole, separated by the scale $s$ and both of finite mass, into a pair of a delocalised, massless 
monopole and its localised, massive antimonopole. The BPS monopole and its antimonopole can be considered static because the attraction mediated by the spatial components of their gauge fields 
(magnetic Coulomb force) is cancelled precisely by the repulsion exerted by the temporal components (Higgs force). In a heroic computation following the lines of 
\cite{'tHooft1976}, where the case of a $|k|=1$ (anti)instanton was investigated, Dmitri Diakonov and 
collaborators have computed the one-loop quantum weight of a nontrivial-holonomy (anti)caloron \cite{Diakonov} for the limit $\frac{s}{\beta}\gg 1$. As it turns out, this limit is the relevant 
one in describing (anti)calorons of the thermal ground state estimate in the deconfining phase of SU(2) Yang-Mills thermodynamics because of the existence of a dimensionless critical temperature $\lambda_c\equiv \frac{2\pi T_c}{\Lambda}=13.87$ and the fact that the emergence of the field $\phi$ is strongly dominated by  (anti)calorons of scale parameter $\rho\sim |\phi|^{-1}$. The result of \cite{Diakonov} shows that such  configurations are unstable under Gaussian quantum noise: large holonomy implies (anti)caloron dissociation into their constituent monopoles 
and antimonopoles whereas small holonomy is driven back to trivial holonomy, 
identifying the latter to be a stable situation.

\subsection{Thermal-ground-state 
estimate: (anti)caloron centers vs. peripheries}
\label{sec2-2}
Having identified $|k|=1$ trivial-holonomy (anti)calorons as viable and relevant field configurations for the a priori estimate of the thermal ground state we now sketch the essential steps in deriving the field $\phi$ from a spatial 
coarse-graining involving the following set of dimensionless phases: 
\eqb
\{\hat{\phi}^a\}\equiv\sum_{C,A}{\rm tr}\!\!\int d^3x \int d\rho \,t^a\,
F_{\mu\nu} (\tau,\mathbf{0}) \,\{(\tau,\mathbf{0}),(\tau,\mathbf{x})\} F_{\mu\nu} (\tau,\mathbf{x}) \{(\tau,\mathbf{x}),
(\tau,\mathbf{0})\}\,,
\label{definition}
\eqe
where 
\begin{equation}
\label{abk}
\{(\tau,\mathbf{0}),(\tau,\mathbf{x})\}\equiv
{\cal P} \exp \left[ i \int_{(\tau,\mathbf{0})}^{(\tau,\mathbf{x})} dz_{\mu} \, A_{\mu}(z) \right]\,,\ \ 
\{(\tau,\mathbf{x}),(\tau,\mathbf{0})\}\equiv\{(\tau,\mathbf{0}),(\tau,\mathbf{x})\}^\dagger\,.
\end{equation}
The Wilson lines in Eq.\,(\ref{abk}) are calculated along
the straight spatial line connecting the points $(\tau,\mathbf{0})$ and $(\tau,\mathbf{x})$, and
${\cal P}$ demands path-ordering. In (\ref{definition}) the sum is over the $|k|=1$ HS 
caloron ($C$) and anticaloron ($A$), and $\{\hat{\phi}^a\}$ signals a family of 
(dimensionless) phases of the field $\phi$ whose continuous parameters 
emerge partially by evaluation of the right-hand side and partially 
relate to temporal shift moduli. It is straight-forward to 
argue \cite{HofmannBook} that (\ref{definition}) is unique: adjointly transforming 
one-point functions as potential integrands vanish identically due to (anti)selfduality, higher $n$-point functions and higher topological charges are excluded by dimensional counting, the coincidence of the spatial (anti)caloron center 
with $\mathbf{0}$ is demanded by spatial isotropy, and the straight-line evaluation of Wilson 
lines is dictated by the absence of any spatial scale on the classical (Euclidean) level. This allows to omit the path-ordering prescription since $A_i$ is a spatial hedge-hog, centered at $\mathbf{0}$, which assigns to each {\sl direction} in ${\bf R}^3$ the same direction in su(2). 

As a result of performing the integrations in Eq.\,(\ref{definition}) one derives $\{\hat{\phi}^a\}$ to be the kernel of the uniquely determined linear differential operator 
${\cal D}\equiv \partial_\tau^2+\left(\frac{2\pi}{\beta}\right)^2$. By absorbing explicit $\beta$ dependence in ${\cal D}\phi=0$ into a potential $V(\phi)$, giving rise to $\phi$'s action density ${\cal L}=\mbox{tr}
\left((\partial_\tau \phi)^2+V(\phi)\right)$, and by demanding consistency between the BPS equations,
\begin{equation}
\label{BPS eqexp}
\partial_\tau\phi=\pm\frac{4\pi i}{\beta} t^3\phi\propto V^{1/2}(\phi)\,,
\end{equation}
and Euler-Lagrange equation 
\begin{equation}
\label{EulLexp}
\partial_\tau^2\phi=\frac{\partial V(\phi^2)}{\partial\phi^2}\phi\,,
\end{equation}
one derives the following first-order equation for $V$ \cite{GiacosaHofmann2007,HofmannBook}
\begin{equation}
\label{eomPot}
\frac{\partial V(|\phi|^2)}{\partial
  |\phi|^2}=-\frac{V(|\phi|^2)}{|\phi|^2}\,.
\end{equation}
Eq.\,(\ref{eomPot}) has the solution 
\begin{equation}
\label{solPot}
V(|\phi|^2)=\frac{\Lambda^6}{|\phi|^2}\,,
\end{equation}
Since the BPS equation for $\phi$ (square root $V^{1/2}$ of the potential $V$ on its right-hand side) needs to be satisfied in addition to the Euler-Lagrange equation (first derivative of $V$ on its right-hand side) the usual additive shift symmetry of the potential due to the Euler-Lagrange equation is not an option: The result $\pm V=\pm 4\pi\Lambda T^3$ (with $|\phi|=\sqrt{\frac{\Lambda^3}{2\pi T}}$) 
turns out to be a good a priori estimate of the thermal ground state's energy density and pressure, respectively, and this result is unique once the integration constant $\Lambda$ has been fixed by experimental data. 

It is important to note that the integration over the instanton scale parameter 
$\rho$ depends {\sl cubically} on the upper integration limit $\rho_u$ 
and that the dependence on $\tau$ of this integral saturates very 
rapidly for $\rho_u/\beta>1$ into a harmonic one. Therefore, the kernel $\{\hat{\phi}^a\}$ 
in Eq.\,(\ref{definition}) of the differential operator ${\cal D}$ is strongly dominated by a small band of $\rho$ values centred at the cutoff $\rho_u=|\phi|^{-1}$, and one can show 
that $\rho_u/\beta=|\phi|^{-1}/\beta\gg 1$ and therefore also $s/\beta\equiv \pi\rho_u^2/\beta^2\gg 1$ for all temperatures of the deconfining phase \cite{HofmannBook}.  

\begin{figure}[h]
\centering
\sidecaption
\includegraphics[width=6cm,clip]{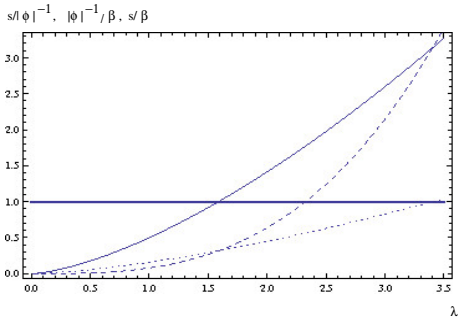}
\caption{Plot of the ratios $\frac{s}{|\phi|^{-1}}$ (solid), $\frac{s}{\beta}$ (dashed), and 
$\frac{|\phi|^{-1}}{\beta}$ (dotted) versus small values of $\lambda$.}
\label{fig-1} 
% Give a unique label
\end{figure}

In discussing the implications of (anti)caloron structure for the physics of the massless mode one neatly needs to distinguish the situation of a thermal ensemble and the consideration of an isolated excitation/disturbance. Let us start by considering an isolated plane wave of mean intensity $I$ and wavelength $l$. For its propagation to be supported by undulating repolarisations of the (anti)selfdual dipole densities provided by trivial-holonomy 
(anti)caloron peripheries, the following inequality needs to be satisfied \cite{HofmannGrandou2015}
\eqb
\label{uncertplaewave}
l\gg \frac{I^2}{8\Lambda^9}\,.
\eqe 
Relation (\ref{uncertplaewave}) is derived from the demand that $l$ needs to be larger 
than length scale $s$ for the caloron periphery to appear like a static and (anti)selfdual 
dipole field,
\eqb
\label{DPF}
l\gg s=\frac12 \lambda^2 \Lambda^{-1}\,,
\eqe
and the fact that $I$ for a plane wave is \cite{HofmannGrandou2015}
\eqb
\label{Ilam}
I=2\lambda \Lambda^4\,.
\eqe
Moreover, one easily derives \cite{HofmannBook} that 
\eqb
\label{sOverPhi-1}
\frac{s}{|\phi|^{-1}}=\frac12\lambda^{3/2}
\eqe 
and 
\eqb
\label{Phi-1overbeta}
\frac{|\phi|^{-1}}{\beta}=\frac{1}{2\pi}\lambda^{3/2}\,.
\eqe 
From Eqs.\,(\ref{sOverPhi-1}), (\ref{Phi-1overbeta}), and from Fig.\,\ref{fig-1} we conclude that the ``dipole length scale" $s$ (the dipole moment) falls 
below the ``coarse-graining length scale" $|\phi|^{-1}$ for sufficiently low temperatures $\lambda$ and that 
$|\phi|^{-1}$ then also falls below $\beta$. Therefore, the derivation of 
$|\phi|^{-1}$, which assumes spatial isotropy and is based on a 
trivial analytic continuation in $\tau$ (staticity of (anti)caloron field strength 
for $r\gg\beta$ leading to saturation of the $\tau$ dependence of (\ref{definition}) for both $r,\rho\gg\beta$) \cite{HerbstHofmann2004}), is not applicable for 
such low temperatures. 
However, relation (\ref{uncertplaewave}) does not depend on $\lambda$, 
and therefore one should regard it as a unique continuation of the high-$\lambda$ 
reasoning to the low-$\lambda$ regime\footnote{The existence of phase boundaries in the thermalised situation is irrelevant when a monochromatic plane wave is considered.}, thus maintaining its validity there. For example, setting $\Lambda\sim 10^4\,$eV and $I\sim 2\times 10^{-4}\,$eV$^4$ (intensity of a 30\,keV synchrotron 
beam of $10^{12}$ photons per second and square millimeter) the requirement on the wavelength $l$ to be in agreement with (\ref{uncertplaewave}) is
\eqb
\label{synchwave}
l\gg 5\times 10^{-51}\,\mbox{m}\,
\eqe
while for $\Lambda\sim 10^{-4}\,$eV one would require $l\gg 5\times 10^{21}\,\mbox{m}$! The situation for a light bulb of intensity $I=500\,$Wm$^{-2}$ requires $l\gg 4.5\times 10^{-53}\,\mbox{m}$ and $l\gg 4.5\times 10^{19}\,\mbox{m}$, respectively, for $\Lambda\sim 10^{4}\,$eV \cite{Hofmann2017} and 
$\Lambda\sim 10^{-4}\,$eV. Therefore, to be in agreement with the observation of wave-like propagation of electromagnetic disturbances up to the X-ray regime, at the same time acknowledging that the bulk of the blackbody spectrum certainly associates with {\sl photonic} fluctuations\footnote{For the present CMB with $T_0=T_c=2.725\,$K and $\Lambda_{\rm CMB}=10^{-4}\,$eV condition (\ref{DPF}), when converted into an upper bound on frequency $\nu(s)$, reads $\nu(s)\ll 220\,$MHz.}, one is led to consider a thermalisation dependent mixing of the Cartan subalgebras in two SU(2) Yang-Mills theories of disparate scales $\Lambda_{\rm CMB}\sim 10^{-4}\,$eV and $\Lambda_{\rm e}\sim 10^{4}\,$eV \cite{Hofmann2017}: Propagating, directed electromagnetic disturbances are waves in SU(2)$_{\rm e}$, a thermal photon gas is described by SU(2)$_{\rm CMB}$ 
with the deep Rayleigh-Jeans regime being subject to wave-like excitations, and intermediates of these 
two extremes are associated with a nontrivial mixing angle.   

\section{SU(2)$_{\rm CMB}$ and the cosmological model}
\label{sec-2}

The postulate that any extended thermal photon gas is subject to SU(2)$_{\rm CMB}$ applies, in particular, to the CMB itself. In this section we will explain how the scale $\Lambda_{\rm CMB}\sim 10^{-4}\,$eV is fixed by the radio excess in the CMB, what SU(2)$_{\rm CMB}$ implies towards the CMB temperature-redshift relation, what this, in turn, means for the dark sector of the cosmological model a high redshifts, and which values cosmological parameters assume in the accordingly modified cosmological model when confronted with temperature and polarisation data as provided in unprecedented signal-to-noise ratios and 
angular resolution by the Planck collaboration.    

\subsection{The postulate of SU(2)$_{\rm CMB}$ describing thermal photon gases}

As already mentioned in the introduction, the critical temperature $T_c$ of 
SU(2)$_{\rm CMB}$ is sharply determined by the present CMB temperature $T_0=2.725\,$K 
\cite{COBE}, resting on the fact that, in cooling the CMB down by virtue of the Universe's expansion, 
the onset of the preconfining phase transition critically generates a Meissner mass $m_\gamma$  
for the formerly massless mode. Frequencies $\nu$ with $\nu<m_\gamma$ and 
$\nu\ll \nu(s)$ then associate with {\sl evanescent} waves whose spectral 
distribution no longer follows the Rayleigh-Jeans law because they do not thermalise anymore \cite{Hofmann2009} and energetically (exponential spatial fall-off of intensity) are 
favoured to occur at zero frequency. In \cite{Hofmann2009} we have, in accord with the 
two conditions above, extracted $m_\gamma\sim 100\,$MHz which is very small 
compared to $T_0=56.7\,$GHz, implying that $T_c$, albeit below $T_0$ is extremely 
close to $T_0$.  

\subsection{Temperature-redshift relation and 3D Ising criticality}

Reviewing the arguments in \cite{HH2018}, let us now discuss one immediate implication of SU(2)$_{\rm CMB}$ with $T_c=T_0$ for the CMB temperature ($T$) - redshift ($z$) relation which is obtained as the solution 
to the equation for energy conservation in an FLRW universe 
\eqb
\label{enecons}
\frac{\mbox{d}\rho}{\mbox{d}a}=-\frac{3}{a}\left(\rho+P\right)\,,
\eqe
where $\rho$, $P$, and $a$ denote SU(2)$_{\rm CMB}$ energy density and pressure, respectively, and $a$ refers to 
the cosmological scale factor, normalized such that today $a(T_0)=1$. Upon appeal to the Legendre transformation between $P$ and $\rho$ and its $T$-differentiated version one finds
\begin{equation}\label{sol}
a = \exp\left(-\frac{1}{3}\int^T_{T_0} \mbox{d}T^\prime \, \frac{\mbox{d}}{\mbox{d}T^\prime} \left[\log \frac{s(T^\prime)}{M^3} \right]\right) = \exp\left(-\frac{1}{3}\log\frac{s(T)}{s(T_0)}\right)\,,
\end{equation}
where $M$ denotes an arbitrary mass scale, and the entropy density $s$ is given as 
\eqb
\label{entropydens}
s=\frac{\rho+P}{T}\,.
\eqe
Note that $s$ does not receive a direct contribution from the thermal ground state \cite{HofmannBook} because such a term cancels in the sum $\rho+P$. The important feature of Eq.\,(\ref{sol}) is 
that its right-hand side solely refers to the boundary values of the state variable $s$ at $T_0$ and $T$. 
In this sense, the solution carries topological information if $s$ at $T_0$ and at $T$ is characterised 
by conformal physics only. This, indeed, is the case for $T_0=T_c$ and $T\gg T_c$ 
(conformal points: $T=T_0$ and $T=\infty$). Namely, for $T\gg T_0$, where the Stefan-Boltzmann limit is well saturated \cite{HofmannBook}, $s(T)$ is proportional to $T^3$. On the other hand, we know that at 
$T_0$ the excitations of the theory represent a free photon gas because the mass of vector modes diverges \cite{HofmannBook}. Therefore, $s(T_0)$ is proportional to $T_0^3$. As a consequence, the ratio $s(T)/s(T_0)$ in Eq.\,(\ref{sol}) reads
\eqb
\label{ratentr}
\frac{s(T)}{s(T_0)}=\frac{g(T)}{g(T_0)} \left(\frac{T}{T_0}\right)^3=\left(\left(\frac{g(T)}{g(T_0)}\right)^{\frac{1}{3}}\frac{T}{T_0}\right)^3\,,\ \ \ (T\gg T_0, T_0=T_c)\,,
\eqe
where $g$ refers to the number of relativistic degrees of freedom at the respective 
temperatures. We have $g(T)=2\times 1+3\times 2=8$ (two photon polarizations plus three polarizations for each of the two vector modes) and $g(T_0)=2\times 1$ (two photon polarizations). Substituting this into Eq.\,(\ref{ratentr}) 
and inserting the result into Eq.\,(\ref{sol}), we arrive at 
\begin{equation}\label{solt>t0}
a = \frac{1}{z+1}=\exp\left(-\frac{1}{3}\log \left(4^{\frac13}\frac{T}{T_0}\right)^3\right) = \left(\frac14\right)^{\frac13}\frac{T_0}{T}\equiv \nu_{\rm CMB}\frac{T_0}{T} \,,\ \ \ (T\gg T_0, T_0=T_c)\,.
\end{equation}
In \cite{HH2017} we have used the 
value $\nu_{\rm CMB}=0.63$ which is a good approximation to 
\eqb
\nu_{\rm CMB}=\left(\frac14\right)^{\frac13}=0.629960(5)\,.
\eqe
In \cite{YaffeSvet} it was argued that the order parameter for deconfinement in 
4D SU(2) Yang-Mills thermodynamics, the 3D Polyakov-loop variable ${\cal P}$, should obey 
long-range dynamics governed by (electric) ${\bf Z}_2$ center symmetry. This is the global symmetry of 
the 3D Ising model void of an external magnetic field. Being members of the same universality class, the critical exponents 
of the correlation lengths thus should coincide in both theories. The superb agreement between 
$\nu_{\rm Ising}=0.629971(4)$ \cite{Kos2016} and $\nu_{\rm CMB}$, 
\eqb       
\left|\frac{\nu_{\rm CMB}-\nu_{\rm Ising}}{\nu_{\rm Ising}}\right|\sim 1.7\times 10^{-5}\,,
\eqe
is indicative of such a link between the two theories even though $\nu_{\rm CMB}$ appears as a 
coefficient in a cosmological $T$-$z$ relation and {\sl not} as a critical exponent 
governing the divergence of the correlation length $l$. To explore this connection further, one may continue the asymptotic behavior of Eq.\,(\ref{solt>t0}) down 
to $T=0$, where $a$ diverges, positing that a strictly monotonic increasing function of the scale factor $a$ describes 
the ratio $l/l_0$ of the Ising correlation length $l$ to some reference length $l_0$ within the critical regime. Note that lowering the temperature through the critical regime at fixed energy content is facilitated by an increase of 
spatial system size ($a$) which implies a much more pronounced increase of the correlation length $l$ as we shall see now. Namely, for the asymptotic, scale-invariant (conformal) solution in (\ref{solt>t0}), continued to $T<T_0$, one may link the fictitious Yang-Mills temperature ratio $T/T_0$ to the physical Ising-model temperature ratio $\tau\equiv\frac{\theta}{\theta_c}$, $\theta_c$ denoting the critical temperature of the Ising phase transition, as
\eqb
\label{IsingYM}
\frac{T}{T_0}=-\frac{1}{\log(\tau-1)}\,.
\eqe 
Obviously, $T\searrow 0$ implies that $\tau\searrow 1$. Substituting Eq.\,(\ref{IsingYM}) into Eq.\,(\ref{solt>t0}) and exponentiating, we arrive 
at
\eqb
\label{Isingcl}
\exp(a)=(\tau-1)^{-\nu_{\rm CMB}}\,.
\eqe 
As $\tau\searrow 1$ this yields the same $l/l_0$ critical behavior for $\exp(a)$, and 
the above-mentioned strictly monotonic function thus turns out to be the exponential map. This means 
that the onset of the divergence of $l/l_0$ can be well discerned at finite system sizes 
as $\tau\searrow 1$. 

\subsection{Dark matter at high redshifts}

In \cite{HH2017,HHK2018} it is explained why the asymptotic solution in Eq.\,(\ref{solt>t0}) 
implies a change of the dark sector in the Hubble parameter $H(z)$ with  
\begin{equation}\label{eq:def:hubbleCosmo}
H^2(z)  = H_0^2 \Big(\Omega_{\rm ds} (z) + \Omega_b (z) + \Omega_{\rm YM} (z) + \Omega_\nu (z) \Big)  \, ,
\end{equation}
where  $\Omega_{b}$ is the baryonic density parameter, $\Omega_{\rm YM}=\rho/\rho_c$ denotes the contribution of the SU(2) plasma, $\rho_c$ is the critical density $\rho_c \equiv 3 H_0^2/(8 \pi G)$, and 
$\Omega_{\nu}$ refers to the neutrino density parameter. The dark-sector density parameter 
$\Omega_{\rm ds} (z)$ exhibits reduced dark matter at high $z$ and interpolates to the 
conventional $\Lambda$CDM model at low $z$:
\begin{equation}
\begin{split}
\label{edmdef}
\Omega_{\rm ds} (z) =\, &\Omega_{\Lambda} + \Omega_{\rm pdm,0} (z+1)^3 + \\ 
&\Omega_{\rm edm,0} \left\{  \begin{array}{lr}
\left(z_{\phantom{p}} + 1\right)^3\,,&  z < z_p\\
\left(z_p + 1\right)^3 \,, & z \geq z_p
\end{array} \right.\,, 
\end{split}
\end{equation}
where the subscripts edm and pdm refer to emergent and primordial dark matter, respectively, 
$\Omega_{\Lambda}$ denotes the ratio to $\rho_c$ of the vacuum energy due to a cosmological constant, and 
$z_p$ is the redshift at which emergent dark matter is released 
instantaneously from a dark-energy like component as the Universe cools down. An important parameter in this  SU(2)$_{\rm CMB}$ driven model of a spatially flat Universe is the fraction $f_p$ of primordial dark matter 
to today's dark matter   
\eqb
f_p\equiv\frac{\Omega_{\rm pdm,0}}{\Omega_{\rm pdm,0}+\Omega_{\rm edm,0}}\,.
\eqe    

\subsection{Results on TT, TE, and EE}

\begin{figure}[h]
\centering
%\sidecaption
\includegraphics[width=10cm,clip]{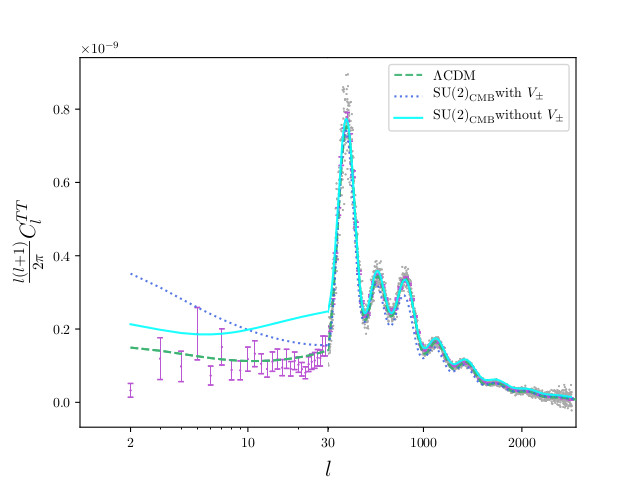}
\caption{The TT power spectra (blue error bars and grey dots are 2015 Planck data, 
dashed line is best-fit $\Lambda$CDM model, dotted line SU(2)$_{\rm CMB}$, also taking $V_\pm$ into account in perturbation equations ($\text{SU(2)}_{\rm CMB}$+$V_\pm$), and solid line SU(2)$_{\rm CMB}$, only taking massless mode into account in perturbation equations ($\text{SU(2)}_{\rm CMB}$).}  
\label{fig-2}       % Give a unique label
\end{figure}

In \cite{HHK2018} the model of Eq.\,(\ref{eq:def:hubbleCosmo}) was fitted to the angular power 
spectra associated with the temperature auto correlation TT, the temperature-E-mode-polarisation cross correlation TE, and the E-mode-polarisation auto correlation based on the 2015 Planck data, see Fig.\,\ref{fig-2} for the case TT. We have used a modified 
CLASS code to compute the spectra and CAMEL to perform the likelihood maximisations 
using lowTEB, HiLLiPOP, and lensing \cite{Aghanim2016}. Leaving out lowTEB does 
not influence the fit results in any essential way.  
Version $\text{SU(2)}_{\rm CMB}$, which is favoured by physical arguments 
on the propagation of temperature perturbations by the wave-like part of the massless mode's spectrum deep in the Rayleigh-Jeans regime \cite{HHK2018}, fits the TT data best (for TE and EE $\text{SU(2)}_{\rm CMB}$+$V_\pm$ and SU(2)$_{\rm CMB}$ can hardly be distinguished) for $l>30$. At low $l$ both versions exhibit excesses. This could be due to the 
omission of (anti)screening effects in the dispersion law of the massless mode 
\cite{LudescherHofmann2008,Falquez2012,Hofmann2013} which imply a dynamical 
breaking of statistical isotropy at low $z$.

\begin{figure}[h]
\centering
\includegraphics[width=14cm,clip]{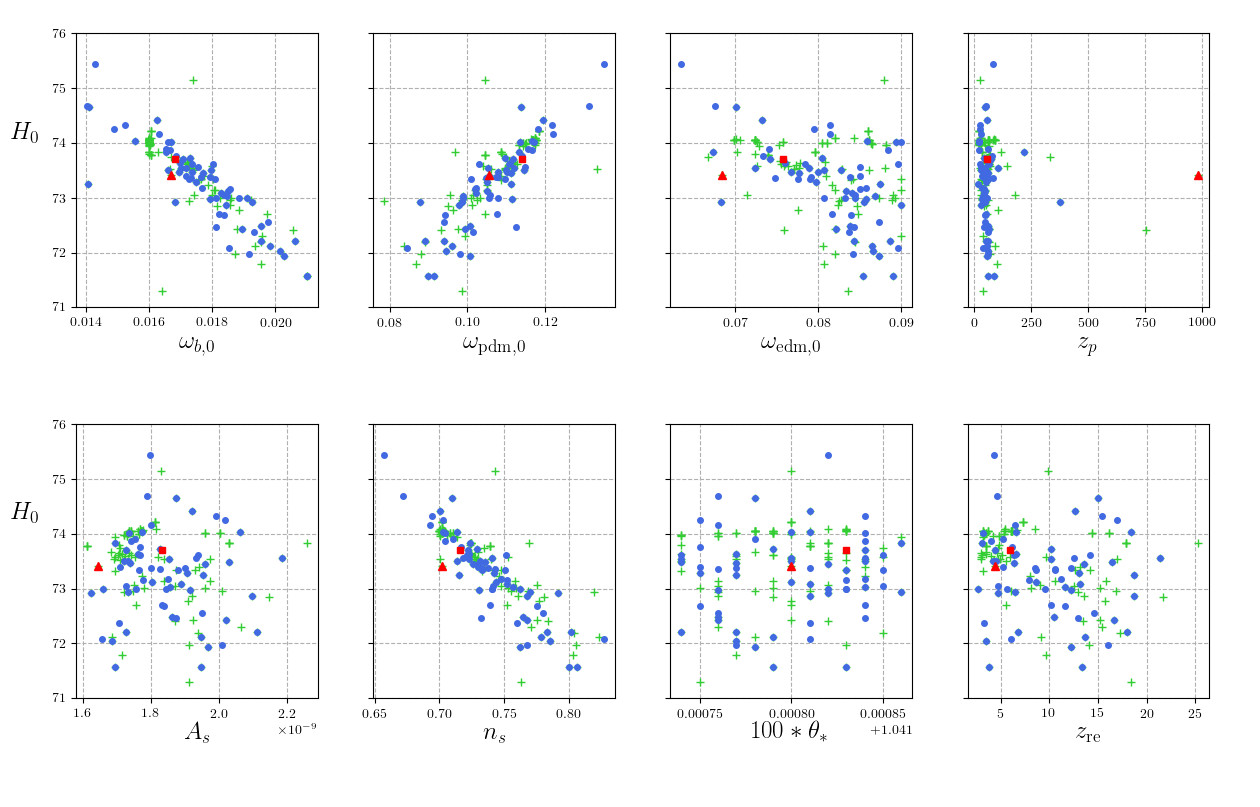}
\caption{Scatter plots of $H_0$ for fitted parameter values, requiring that $\chi^2<21700$ (best-fit: $\chi^2=21192.6$ for $\text{SU(2)}_{\rm CMB}$, $\chi^2=20940.1$ for $\text{SU(2)}_{\rm CMB}$+$V_\pm$). Crosses and dots refer to $\text{SU(2)}_{\rm CMB}$+$V_\pm$ and $\text{SU(2)}_{\rm CMB}$, whose best-fit parameter values are indicated by a solid triangle and a solid square, respectively.}  
\label{fig-3}
\end{figure}

In Fig.\,\ref{fig-3} the value of $H_0$ versus other key cosmological 
parameters is depicted for likelihood maximisations, which 
do not exceed $\chi^2=21700$, to characterise their scatter. 
Notably, we obtain a low value for the re-ionisation redshift $r_{\rm re}$, essentially in agreement 
with its extraction in detecting the Gunn-Peterson trough in high-redshift quasar spectra \cite{Becker}, a 
low value of the baryon density which deviates by about 30\% from the BBN deuterium concordance  
value but matches the value of typical baryon censuses \cite{Shull}, see \cite{Nicastro2018}, however, for an apparent, alternative resolution of the missing-baryon problem, and a best-fit value of $H_0$ coincident with the  
most recent and most precise SNe Ia distance-redshift extraction \cite{Riess2018}. Notice the low value of $n_s$, demanding a red-tilted spectrum of primordial, adiabatic curvature perturbations.       

\section{Summary and Outlook}
\label{sec-3}

In this talk we have reviewed the construction of the thermal ground state in the deconfining phase of SU(2) Yang-Mills thermodynamics. The outcome of this construction imposes a strong contraint on 
the (anti)caloron scale parameter, $\rho=|\phi|^{-1}$, and therefore on the static configuration of the (anti)selfdual 
field outside the central region where spatial coarse-graining is performed. A relation between a classical plane wave's 
intensity $I$, its frequency $\nu$, and the Yang-Mills scale $\Lambda$ of the SU(2) theory is obtained by appeal 
to the condition that the wavelength $l$ must be larger than the dipole scale $s=\pi\frac{\rho^2}{\beta}=\frac12 \Lambda^{-1}\lambda^2$. Because $s$ and $I$ do depend on $\lambda$ this relation does not exhibit any dependence on $\lambda$ and therefore can be 
(trivially) continued to small values of $\lambda$, permitting very small wavelengths to classically propagate provided that $\Lambda$ is sufficiently large. To cover Max Planck's spectral intensity of a blackbody for {\sl all} thermal photon gases and the propagation of isolated electromagnetic waves of so-far measured/observed intensity and frequency,  at least two SU(2) Yang-Mills theories of 
disparate scales $\Lambda_{\rm CMB}\sim 10^{-4}\,$eV \cite{Hofmann2009} and $\Lambda_{\rm e}\sim 10^{4}\,$eV \cite{Hofmann2017} -- SU(2)$_{\rm CMB}$ and SU(2)$_{\rm e}$ -- are thus required, subject to a thermalisation dependent mixing angle for their Cartan subalgebras. Subsequently, implications of SU(2)$_{\rm CMB}$ for the cosmological model were elucidated. In a first step, the $T$-$z$ relation for the CMB, which exhibits curvature at low $z$ and asymptotes into a modified linear behaviour at 
large $z$, was reviewed. The asymptotic coefficient in this relation is strongly conjectured to be linked to the critical exponent for the correlation length in the 3D Ising transition in the absence of an external magnetic field, and we 
have given arguments why such a link should exist. Finally, we have explored the implications 
of the new $T$-$z$ relation for the dark-matter sector of the flat-space FLRW cosmological model. Fits to the 2015 Planck data on TT, TE, and EE have revealed values for directly measurable parameters ($H_0$, $z_{\rm re}$, and $\omega_{b,0}$) that are in line with {\sl local} cosmological observations but exhibit considerable tension with global fits to the standard cosmological model (including BBN and a high-$z$ confidence in $\Lambda$CDM). 

An important target of future investigation is the effect of SU(2)-induced (anti)screening in the dispersion law of the massless mode \cite{LudescherHofmann2008,Falquez2012,Hofmann2013}. Since this effect breaks statistical isotropy dynamically 
one should simulate an ensemble of CMB skies, search for matches of angular power spectra, and consider appropriately designed statistics \cite{Schwarz} to assess whether typical members are extremely unlikely in the standard $\Lambda$CDM model.

\end{document}